\documentclass{article}

\def\xxinput#1{\input#1}

\xxinput{vsolj02.sty}

\usepackage[dvipdfmx]{graphicx}
\usepackage[comma,colon]{natbib}

\usepackage[OT2,T1]{fontenc}

\def\cite{\citealt}
\setcitestyle{aysep={}}

\newcounter{author}
\def\altaffilmark#1{$^{#1}$}
\def\altaffiltext#1{$^{#1}$\,}

\def\authorcount#1#2{{\refstepcounter{author}\label{#1}
                     \altaffiltext{\ref{#1}}{#2}}}

\begin{document}

\begin{center}

\title{SDSS J183131.63$+$420220.2: AM CVn star showing}
\vskip -2mm
\title{ER UMa-type behavior and long standstill}

\author{
        Taichi~Kato\altaffilmark{\ref{affil:Kyoto}}}
\email{tkato@kusastro.kyoto-u.ac.jp}

\authorcount{affil:Kyoto}{
     Department of Astronomy, Kyoto University, Sakyo-ku,
     Kyoto 606-8502, Japan}

\end{center}

\begin{abstract}
\xxinput{abst.inc}
\end{abstract}

   SDSS J183131.63$+$420220.2 was selected as a white dwarf
candidate from Sloan Digital Sky Survey (SDSS) colors and was
classified as a cataclysmic variable (CV) by \citet{gir11SDSSWDs}.
This object was spectroscopically classified as a DB white dwarf
by \citet{kle13SDSSWDs}.
The variability of this object was detected by Asteroid
Terrestrial-impact Last Alert System (ATLAS: \cite{ATLAS})
and was listed as ATO J277.8818$+$42.0389 with a classification
of ``irregular'' \citep{hei18ATLASvar}.\footnote{
   Although I used this as the primary name in vsnet-alert 27896
($<$http://ooruri.kusastro.kyoto-u.ac.jp/mailarchive/vsnet-alert/27896$>$),
   which first described variability, I here use the SDSS name knowing
   that the CV nature had already been recognized.
}  The variation was also detected by Zwicky Transient Facility
(ZTF: \cite{ZTF}) and was given a name of ZTF J183131.63$+$420220.1
\citep{ofe20ZTFvar}.  Most recently, \citet{GaiaDR3} listed
this object as a CV (Gaia DR3 2111270034246759424).

   \citet{ini23SDSSCVs} studied this object and classified it
as an AM CVn star with a helium-dominated absorption line
spectrum and showing occasional ``drop-outs'' (fading)
in the ZTF and Gaia light curves.  \citet{ini23SDSSCVs} described
that the drop-outs in the Gaia and ZTF light curves are
unusual for this type (i.e. AM CVn) of system.

\begin{figure*}
\begin{center}
\includegraphics[width=16cm]{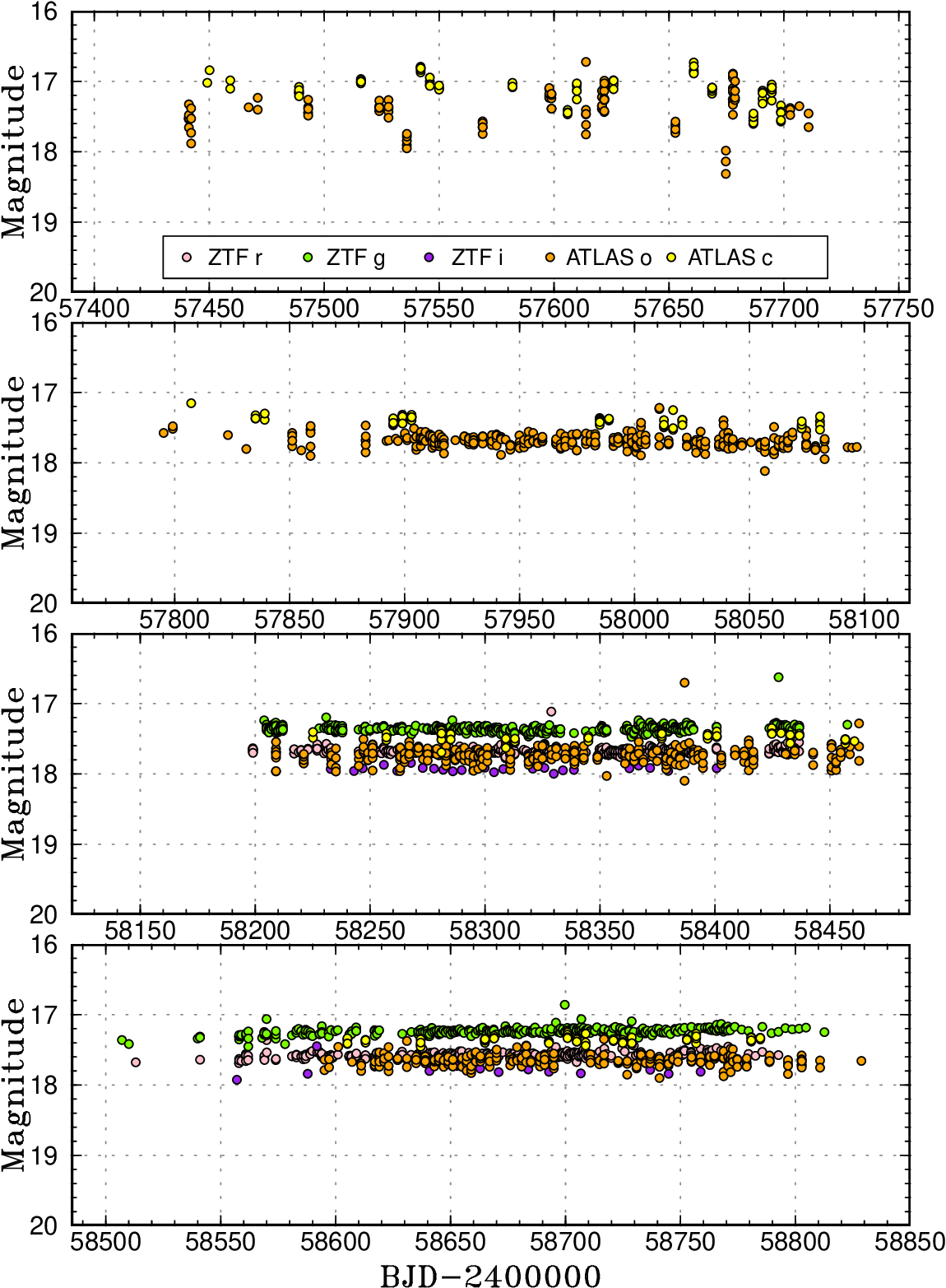}
\caption{
   Light curve of SDSS J183131.63$+$420220.2 in 2016--2019.
   The object was mostly in nearly constant brightness in
   2017--2019.  Larger-amplitude variations were present in
   2016, as was also evident in Gaia observations shown in
   \citet{ini23SDSSCVs}.
}
\label{fig:lc1}
\end{center}
\end{figure*}

\begin{figure*}
\begin{center}
\includegraphics[width=16cm]{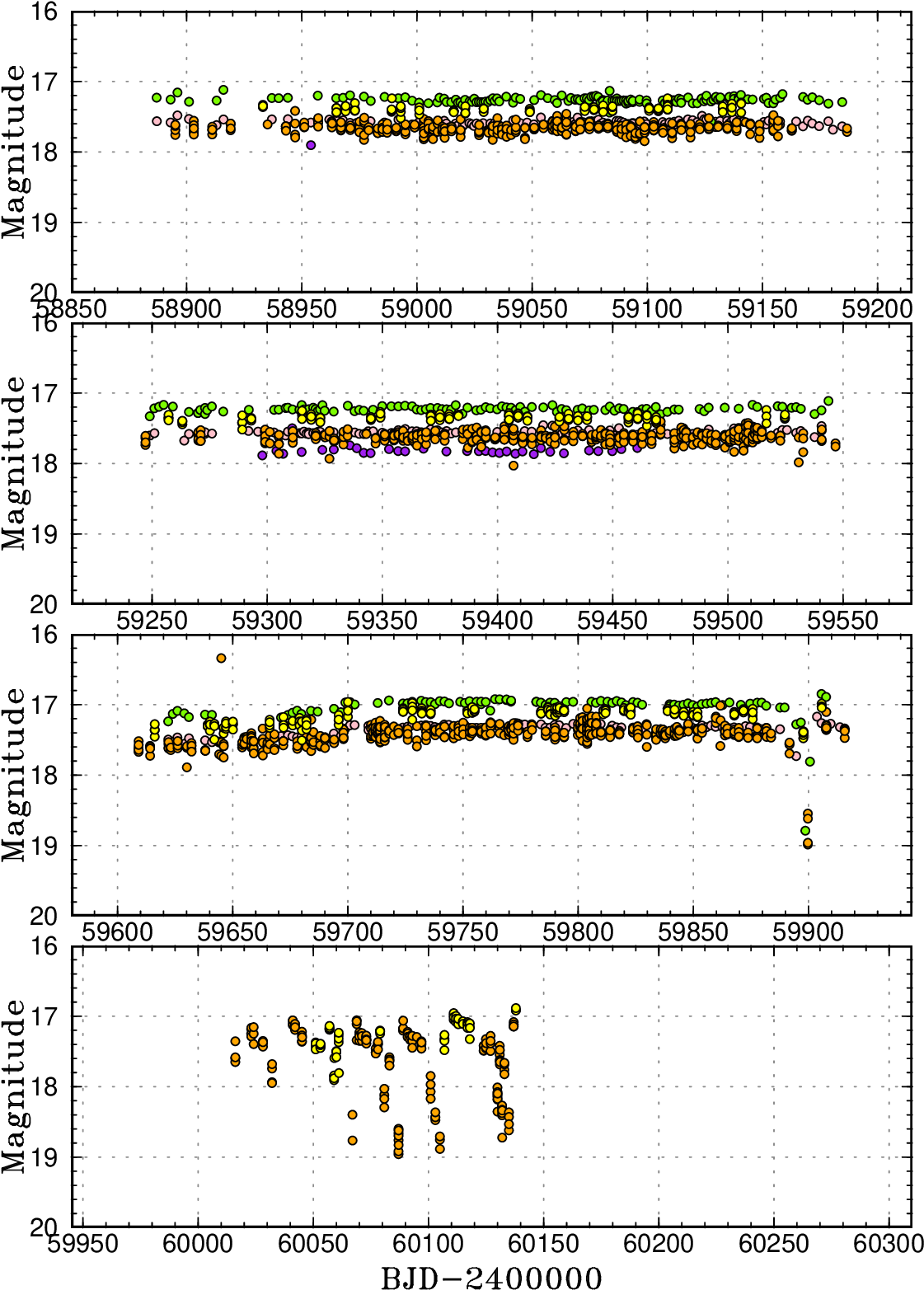}
\caption{
   Light curve of SDSS J183131.63$+$420220.2 in 2020--2023.
   The symbols are the same as in figure \ref{fig:lc1}.
   The object brightened somewhat in 2022 (third panel)
   with a sudden short drop near the end.  The object
   was in dwarf nova (ER UMa) phase in 2023 (last panel).
}
\label{fig:lc2}
\end{center}
\end{figure*}

\begin{figure*}
\begin{center}
\includegraphics[width=16cm]{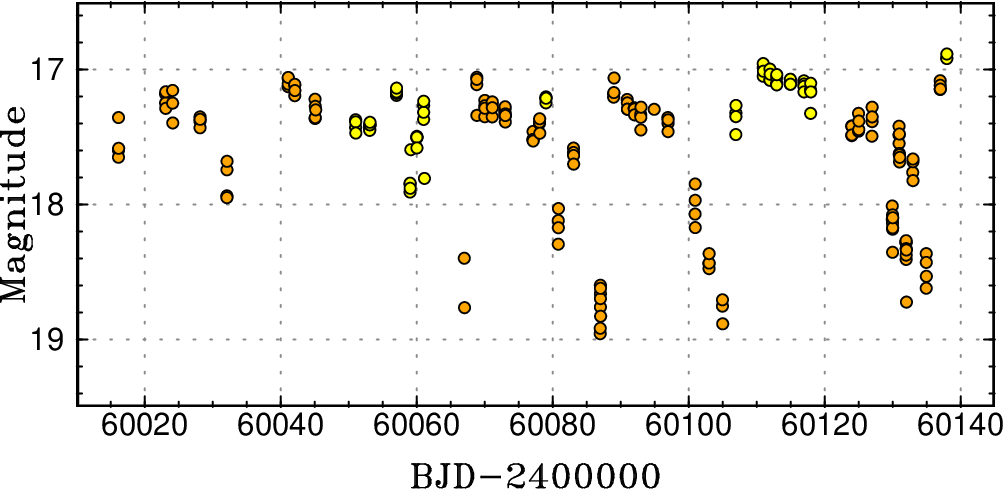}
\caption{
   Enlargement of the light curve of SDSS J183131.63$+$420220.2
   in 2023.  The symbols are the same as in figure \ref{fig:lc1}.
   Long outbursts were intervened by short fading episodes.
   Short outbursts are also also evident around BJD 2460131 and
   2460133.
}
\label{fig:lc2023}
\end{center}
\end{figure*}

   Using ZTF and ATLAS data, I found that this object currently
(in 2023) shows ER UMa-type
\citep[subclass of SU UMa stars; see e.g.][]{kat99erumareview}
variations (vsnet-alert 27896).  The long-term light curves
are shown in figures \ref{fig:lc1} and \ref{fig:lc2}.
The enlarged 2023 light curve is shown in figure \ref{fig:lc2023}.
Long outbursts (superoutbursts) were intervened by short fading episodes.
Short outbursts are also also evident around BJD 2460131 and
2460133.  The recurrence times of long outbursts varied
between 20 and 30~d.  The duty cycles of long outbursts
were $\sim$0.5 or even larger.  The current light curve is
very similar to the AM CVn star MGAB-V240 particularly
in 2021 to 2022 \citep{kat23mgabv240}.  It is also similar
to the hydrogen-rich ER UMa star RZ LMi when it showed
standstills \citep{kat16rzlmi}.  MGAB-V240 was the second
AM CVn-type object showing true standstills after
CR Boo \citep{kat23crboo}.  SDSS J183131.63$+$420220.2 has become
the third known such an object and is notable in that
it showed a very long standstill lasting almost 6~yr,
while the other two objects showed relatively short ones.
The suggested type is SU UMa(ER UMa)+Z Cam+AM CVn,
the same as for MGAB-V240.

\begin{figure*}
\begin{center}
\includegraphics[width=14cm]{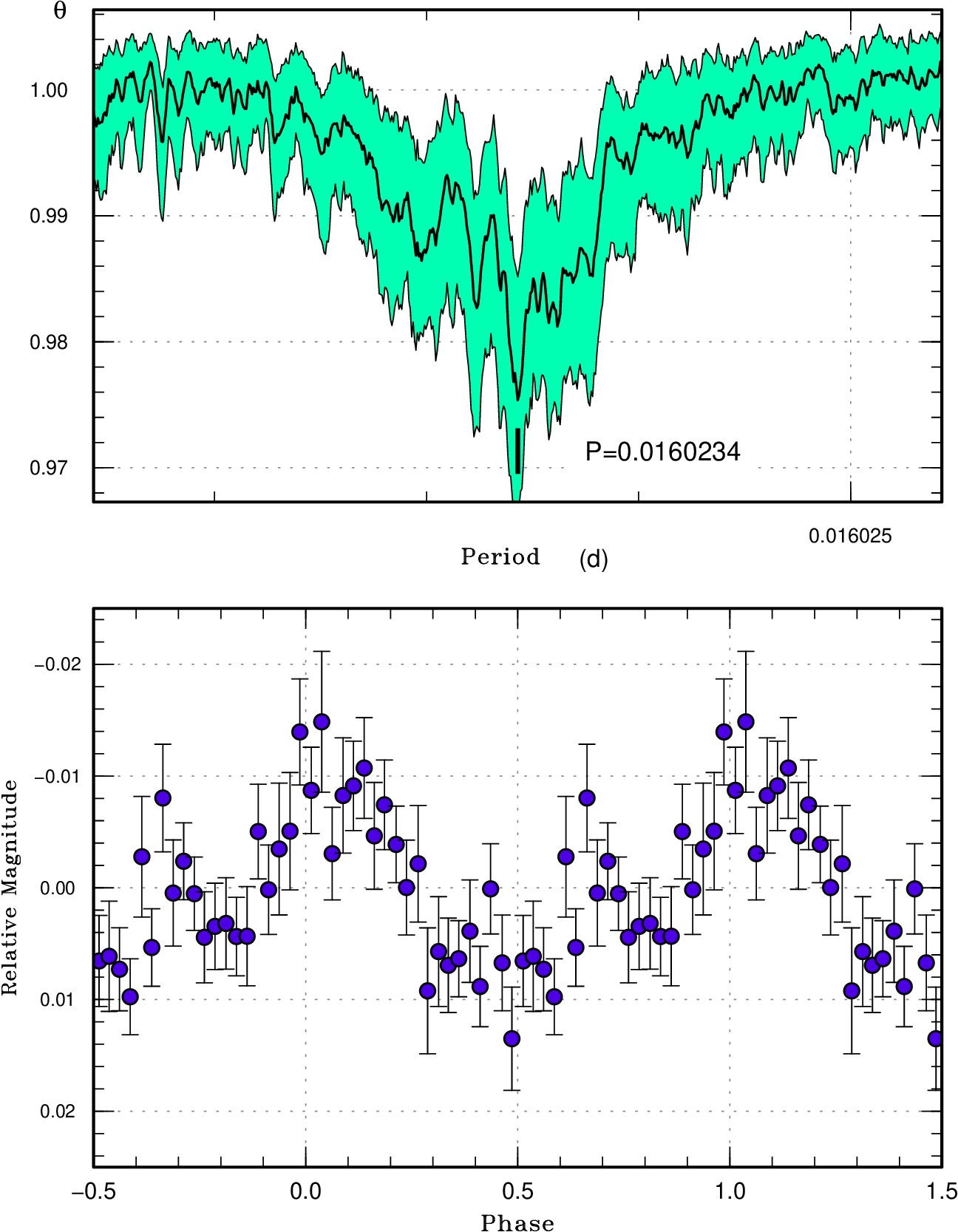}
\caption{
   Period analysis of SDSS J183131.63$+$420220.2 during
   the standstill.
   (Upper): PDM analysis.  The bootstrap result using
   randomly contain 50\% of observations is shown as
   a form of 90\% confidence intervals in the resultant 
   $\theta$ statistics.
   (Lower): Phase plot.
}
\label{fig:pdm}
\end{center}
\end{figure*}

   The discussion in MGAB-V240 \citep{kat23mgabv240} also
applies to SDSS J183131.63$+$420220.2.  The accretion disk
in this object should be very close to the thermal stability.
Using the ZTF data during the standstill, I detected
a period of 0.01602343(1)~d (figure \ref{fig:pdm}) using
the phase dispersion minimization (PDM: \cite{PDM}) method
after removing long-term trends by using locally-weighted
polynomial regression (LOWESS: \cite{LOWESS}).
The error was estimated by the methods of \citet{fer89error,Pdot2}.
There was no other candidate period in the period region
acceptable for an AM CVn star.  This period is likely
either the orbital or superhump period of this system.
I consider the former more likely based on the long coherence.
If the period is the orbital period, a superhump signal
$\sim$1\% different from this can be expected.  I could not
find such a signal.  The period in SDSS J183131.63$+$420220.2
is very close to the superhump period [0.015824(9)~d]
of MGAB-V240.  It is most likely that the limit of
the thermal stability in the disks of AM CVn stars is around
these periods, as expected from the disk-instability theory
\citep{tsu97amcvn,sol10amcvnreview,kot12amcvnoutburst}.

   The long-term variation during the standstill probably
requires long-term slight variation of the mass-transfer rate,
as is usually considered as the origin of standstills in
hydrogen-rich Z Cam stars \citep{mey83zcam}.
I also point out that the gradually rising standstill
(2017 to 2021) and brightening in late 2022 before its termination
are somewhat similar to the IW And stars (hydrogen-rich systems)
\citep{szk13iwandv513cas,kat19iwandtype}.  The time scales
in SDSS J183131.63$+$420220.2, however, would be too long
compared to the viscous time scale of the disk to be considered
as the same origin as in IW And stars.

\section*{Acknowledgements}

This work was supported by JSPS KAKENHI Grant Number 21K03616.

I am grateful to Naoto Kojiguchi for helping downloading the ZTF data
and the ATLAS and ZTF teams for making their data
available to the public.

This work has made use of data from the Asteroid Terrestrial-impact
Last Alert System (ATLAS) project.
The ATLAS project is primarily funded to search for
near earth asteroids through NASA grants NN12AR55G, 80NSSC18K0284,
and 80NSSC18K1575; byproducts of the NEO search include images and
catalogs from the survey area. This work was partially funded by
Kepler/K2 grant J1944/80NSSC19K0112 and HST GO-15889, and STFC
grants ST/T000198/1 and ST/S006109/1. The ATLAS science products
have been made possible through the contributions of the University
of Hawaii Institute for Astronomy, the Queen's University Belfast, 
the Space Telescope Science Institute, the South African Astronomical
Observatory, and The Millennium Institute of Astrophysics (MAS), Chile.

Based on observations obtained with the Samuel Oschin 48-inch
Telescope at the Palomar Observatory as part of
the Zwicky Transient Facility project. ZTF is supported by
the National Science Foundation under Grant No. AST-1440341
and a collaboration including Caltech, IPAC, 
the Weizmann Institute for Science, the Oskar Klein Center
at Stockholm University, the University of Maryland,
the University of Washington, Deutsches Elektronen-Synchrotron
and Humboldt University, Los Alamos National Laboratories, 
the TANGO Consortium of Taiwan, the University of 
Wisconsin at Milwaukee, and Lawrence Berkeley National Laboratories.
Operations are conducted by COO, IPAC, and UW.

The ztfquery code was funded by the European Research Council
(ERC) under the European Union's Horizon 2020 research and 
innovation programme (grant agreement n$^{\circ}$759194
-- USNAC, PI: Rigault).

\section*{List of objects in this paper}
\xxinput{objlist.inc}

We provide two forms of the references section (for ADS
and as published) so that the references can be easily
incorporated into ADS.

\newcommand{\noop}[1]{}\newcommand{\hyphalt}{-}

\renewcommand\refname{\textbf{References (for ADS)}}

\xxinput{j183131aph.bbl}

\renewcommand\refname{\textbf{References (as published)}}

\xxinput{j183131.bbl.vsolj}

\end{document}